\documentclass[conference]{IEEEtran}
\IEEEoverridecommandlockouts
\usepackage{cite}
\usepackage{amsmath,amssymb,amsfonts}
\usepackage{algorithmic}
\usepackage{graphicx}
\usepackage{textcomp}
\usepackage{xcolor}
\usepackage{stfloats}
\usepackage{booktabs}
\usepackage{subcaption}
\usepackage[font=footnotesize]{caption}
\usepackage{pbalance}

\usepackage{hyperref}
\hypersetup{
	colorlinks=true,
	linkcolor=black,
	citecolor=black,
	urlcolor=black
}

\def\BibTeX{{\rm B\kern-.05em{\sc i\kern-.025em b}\kern-.08em
    T\kern-.1667em\lower.7ex\hbox{E}\kern-.125emX}}
\begin{document}

\title{Real-Time Co-Simulation for DC Microgrid Energy Management with Communication Delays}

\author{
\IEEEauthorblockN{\textbf{S. Gokul Krishnan, Mohd Asim Aftab, Shehab Ahmed, and Charalambos Konstantinou}}

\IEEEauthorblockA{
CEMSE Division, King Abdullah University of Science and Technology (KAUST)}
\IEEEauthorblockA{E-mail: \{gokulkrishnan.sivakumar, mohammad.aftab, \\ shehab.ahmed, charalambos.konstantinou\}@kaust.edu.sa}
}

\maketitle
\begin{abstract}
The growing integration of renewable energy sources (RESs) in modern power systems has intensified the need for resilient and efficient microgrid solutions. DC microgrids have gained prominence due to their reduced conversion losses, simplified interfacing with DC-based RESs, and improved reliability. To manage the inherent variability of RESs and ensure stable operation, energy management systems (EMS) have become essential. While various EMS algorithms have been proposed and validated using real-time simulation platforms, most assume ideal communication conditions or rely on simplified network models, overlooking the impact of realistic communication delays on EMS performance. This paper presents a novel real-time cyber-physical system (CPS) testbed for evaluating EMS performance in DC microgrids under realistic communication delays. The proposed testbed integrates a DC microgrid modeled in OPAL-RT with an EMS controller implemented on a Raspberry Pi (RPi). The communication network is emulated using EXataCPS, enabling the exchange of actual power system traffic and the replication of realistic latency conditions. This comprehensive setup captures the interplay between power system dynamics, EMS control, and communication network behavior.
\end{abstract}


\section{Introduction} \label{sec: intro}
The rapid integration of renewable energy sources (RESs) into modern power systems has underscored the critical need for resilient microgrid architectures capable of maintaining stability, efficiency, and grid compliance \cite{DER_Review}. Among these, DC microgrids have emerged as a promising solution due to their inherent advantages in reducing conversion losses, simplifying integration with DC-based RES like photovoltaics (PV), and enhancing reliability compared to AC counterparts \cite{DCuG_Review}. With the increasing integration of RESs and rising consumer demand, the implementation of an effective energy management system (EMS) in DC microgrids has become essential \cite{EMSinMG, CMS_DC, mohammed2024decentralized}. EMS addresses challenges such as the stochastic nature of PV output and battery utilization by optimizing the use of available resources \cite{EMS_advanatgeforRES1}, \cite{EMS_advantageforRES}. 

EMS leveraging optimization algorithms have been explored to coordinate battery energy storage systems (BESS), diesel generators, and load curtailment, ensuring power balance during islanding transitions \cite{CentralizedEMS}. 
Authors in \cite{SoC_DC} propose a state-of-charge (SoC) based EMS to enhance DC bus stability, battery lifespan, and transient performance in DC microgrid. To effectively manage multiple energy storage systems, multi-level energy management has been proposed in literature \cite{EMS_DC}.

The successful validation and deployment of EMS in microgrids requires comprehensive testing on real-time CPS testbeds that accurately replicate both the electrical infrastructure and the communication network. This integration is essential, as the reliability, responsiveness, and overall performance of EMS solutions critically depend on real-time data exchange enabled by the underlying communication architecture. While prior works \cite{CMS_DC}, \cite{CentralizedEMS} have demonstrated EMS validation on real-time platforms, they often overlook the role of the communication network. Authors in \cite{RealTimeEMS} incorporate fixed communication delays to examine their impact on EMS performance, while \cite{OPF_undercommdelay} models communication latency as a stochastic process. To ensure robust, practical, and deployable EMS solutions, it is imperative to incorporate a complete real-time CPS  testbed that accurately emulates both power system dynamics and realistic communication network conditions. Such holistic validation is essential to identify potential performance bottlenecks and cyber-physical interactions before field deployment, ultimately advancing the reliability and efficiency of EMS.

To bridge this research gap, this paper presents a novel real-time cyber-physical system (CPS) testbed for evaluating EMS performance in DC microgrids under realistic communication delays. The DC microgrid model is executed in OPAL-RT, while the EMS controller is deployed on a control hardware-in-the-loop (CHIL) setup using a Raspberry Pi (RPi). The communication network is accurately emulated in EXataCPS, which serves as a real-time interface between the OPAL-RT power system and the EMS controller. The developed testbed offers a novel platform for validating EMS solutions in DC microgrids by utilizing real power system traffic over the communication network. By incorporating real-time system signals, it enables precise and realistic evaluation of EMS control strategies under operational conditions. By bridging the gap between physical simulation, algorithmic optimization, and network emulation, this work advances the development of resilient microgrid architectures capable of operating in complex, delay-prone environments.

The rest of the paper is organized as follows. Section \ref{sec: EMS} presents modeling of EMS for DC microgrid. Section \ref{sec:testbed} presents a detailed description of the CPS testbed for evaluating EMS under realistic communication delays. Section \ref{sec:results} presents the results of EMS on the developed testbed. Finally, Section \ref{sec:conclusion} concludes the paper.

\section{Energy Management System (EMS)}\label{sec: EMS}

The EMS serves as the central decision-making engine that couples the interactions among PV generation, BESS, and the grid. Its primary objective is to minimize the total operational cost while satisfying power-balance, SoC constraints, and system reliability requirements. The EMS operates on a receding‐horizon basis, incorporating real‐time measurements to generate optimal control actions for each resource over a finite time window. By continually updating its plan in response to changing load, generation, and price signals, the EMS ensures both economic efficiency and operational robustness.



The optimization strategy models the operation of each battery \(i\in\{1,\dots,n\}\) at every hour \(t\in\{0,\dots,23\}\) using a single integer decision variable per time step given as (\ref{eq:batcommand}): 
\begin{equation}
    d_{i,t}\;\in\;\{-1,\,0,\,1\}
    \label{eq:batcommand}
\end{equation}
\(d_{i,t}=+1\) directs battery \(i\) to discharge at its full power rating \(P^{\max}_{i}\), \(d_{i,t}=-1\) initiates to charge at the same rate, and \(d_{i,t}=0\) leaves the battery idle. By unifying the charge, idle, and discharge commands into a single integer variable, the EMS reduces computational complexity and facilitates real-time implementation on embedded platforms.


Once \(d_{i,t}\) is selected, the actual power exchanged by battery \(i\) is given by (\ref{eq:power}): 
\begin{equation}
P_{b,i,t} \;=\; d_{i,t}\,P^{\max}_{i}
\label{eq:power}
\end{equation}
where the $P^{\max}_{i}$ is the max charge/discharge rate of the battery $i$, with positive values indicating discharge into the network and negative values indicating charging.  The net power exchanged with the main grid, \(P_{g,t}\), is then determined by enforcing instantaneous power balance across all $n$ buses, as in (\ref{eq:netpower}):
\begin{equation}
P_{g,t}
\;=\;\sum_{i=1}^{n}P^{\mathrm{load}}_{i,t}
\;-\;\sum_{i=1}^{n}P^{\mathrm{PV}}_{i,t}
\;-\;\sum_{i=1}^{n}P_{b,i,t}.
\label{eq:netpower}
\end{equation}
where \(P^{\mathrm{PV}}_{i,t}\) denotes the PV generation at bus \(i\), and \(P^{\mathrm{load}}_{i,t}\) denotes the local demand.  A positive \(P_{g,t}\) means net import from the grid, while a negative value corresponds to export.

The stored energy \(E_{i,t}\) in battery \(i\) is updated according to the SoC dynamics and is given by (\ref{eq:storedenergy}): 
\begin{equation}
E_{i,t+1}
\;=\;
E_{i,t}
\;+\;\eta\,P_{b,i,t}\,\Delta t,
\label{eq:storedenergy}
\end{equation}
where \(\eta\) is the round‐trip efficiency and \(\Delta t\) is the one‐hour time step.  A discharge action \((d_{i,t}=+1)\) reduces \(E_{i,t+1}\) by \(\eta\,P^{\max}_i\), whereas a charge action \((d_{i,t}=-1)\) increases it by the same amount.

To maintain battery health, the stored energy is constrained at each time to lie within safe bounds: 
\begin{equation}
\underline{\mathrm{SoC}}_{i}\,C_{i}
\;\le\;
E_{i,t}
\;\le\;
\overline{\mathrm{SoC}}_{i}\,C_{i},
\label{eq:bounds}
\end{equation}
where \(C_{i}\) is the energy capacity of battery \(i\) (kWh), and \(\underline{\mathrm{SoC}}_{i}\), \(\overline{\mathrm{SoC}}_{i}\) are the minimum and maximum permissible state‐of‐charge fractions, respectively.   

Finally, the EMS minimizes the total economic cost over the horizon, combining the cost of grid transactions and the revenue from battery discharge. Its objective function is shown in (\ref{eq:objfn}): 
\begin{equation}
min \sum_{t=0}^{T-1}\Bigl(C^{\mathrm{grid}}_{t}\,P_{g,t}
\;-\;\sum_{i=1}^{n}C^{\mathrm{bess}}_{t}\,P_{b,i,t}\Bigr),
\label{eq:objfn}
\end{equation}
where \(C^{\mathrm{grid}}_{t}\) is the grid electricity price and \(C^{\mathrm{bess}}_{t}\) is the per‐kWh remuneration for battery discharge at hour \(t\).  This discrete‐control approach yields a clear, implementable rule set that charges when prices are low, discharges when high, and idles otherwise while rigorously respecting power balance, SoC dynamics, and operational limits at all buses.

\section{Real-Time Cyber-Physical System (CPS) Testbed Development}\label{sec:testbed}
\begin{figure}[t]
    \centering
    \includegraphics[scale=0.5]{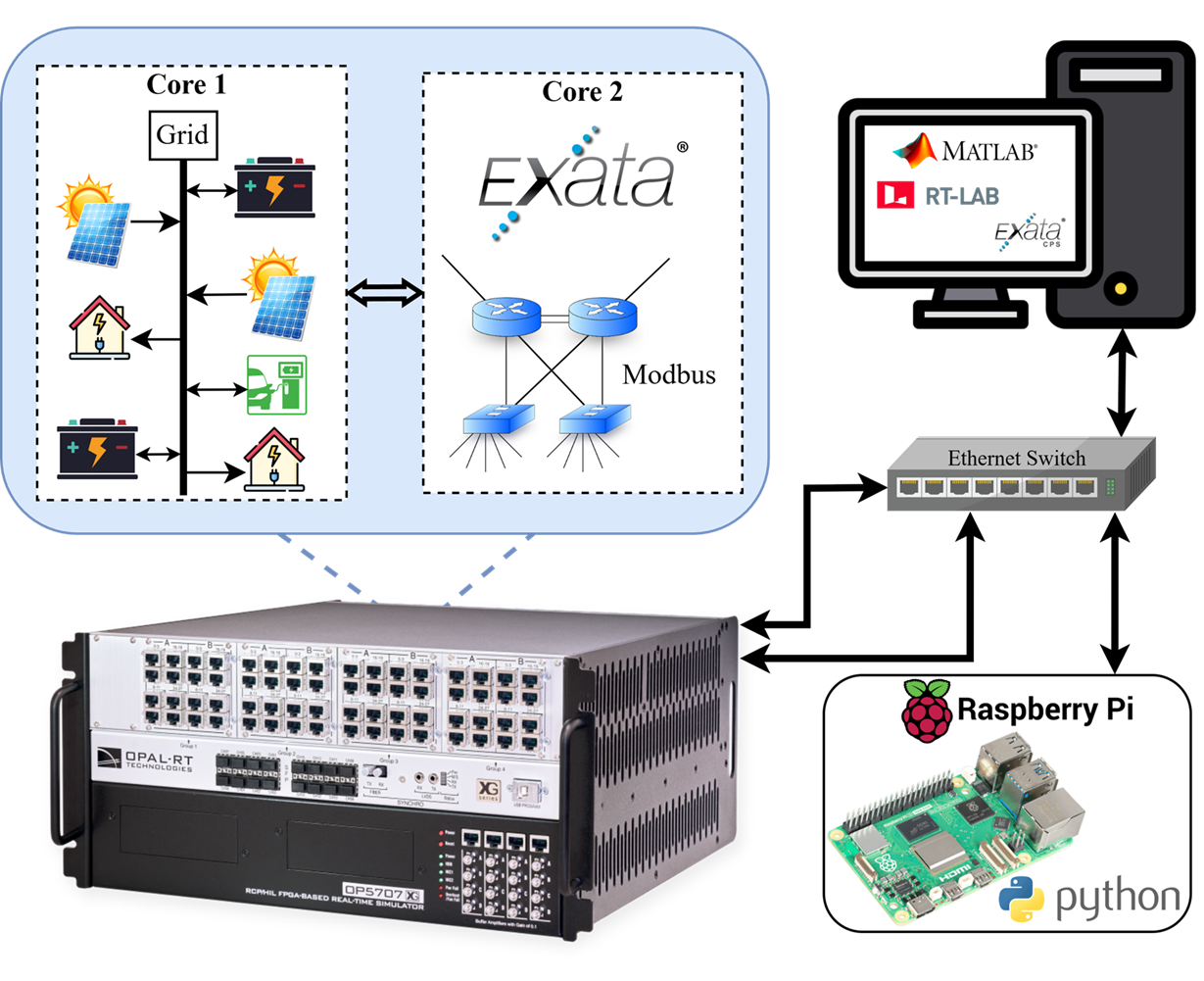}
    \caption{Overview of the real-time CPS testbed architecture.}
    \vspace{-3mm}
    \label{fig:testbed}
\end{figure}

The real-time CPS testbed consists of three integrated components: a grid-integrated DC microgrid executed in OPAL-RT using MATLAB Simulink and RT-LAB, a communication network simulation modeled using EXataCPS, and an optimization-based EMS implemented in Python as shown in Fig. \ref{fig:testbed}. The CPS testbed architecture is structured into three interconnected layers. The physical layer, running on one core of the OPAL-RT simulator, hosts the real-time simulation of the DC microgrid. The communication layer, executed on a separate OPAL-RT core, simulates the communication network using EXataCPS. The control layer is deployed on an RPi, which is physically connected to the simulated network via an Ethernet switch and runs the Python-based EMS in real-time.

\subsection{System Modeling}

\begin{figure}[t]
    \centering
    \includegraphics[scale=0.07]{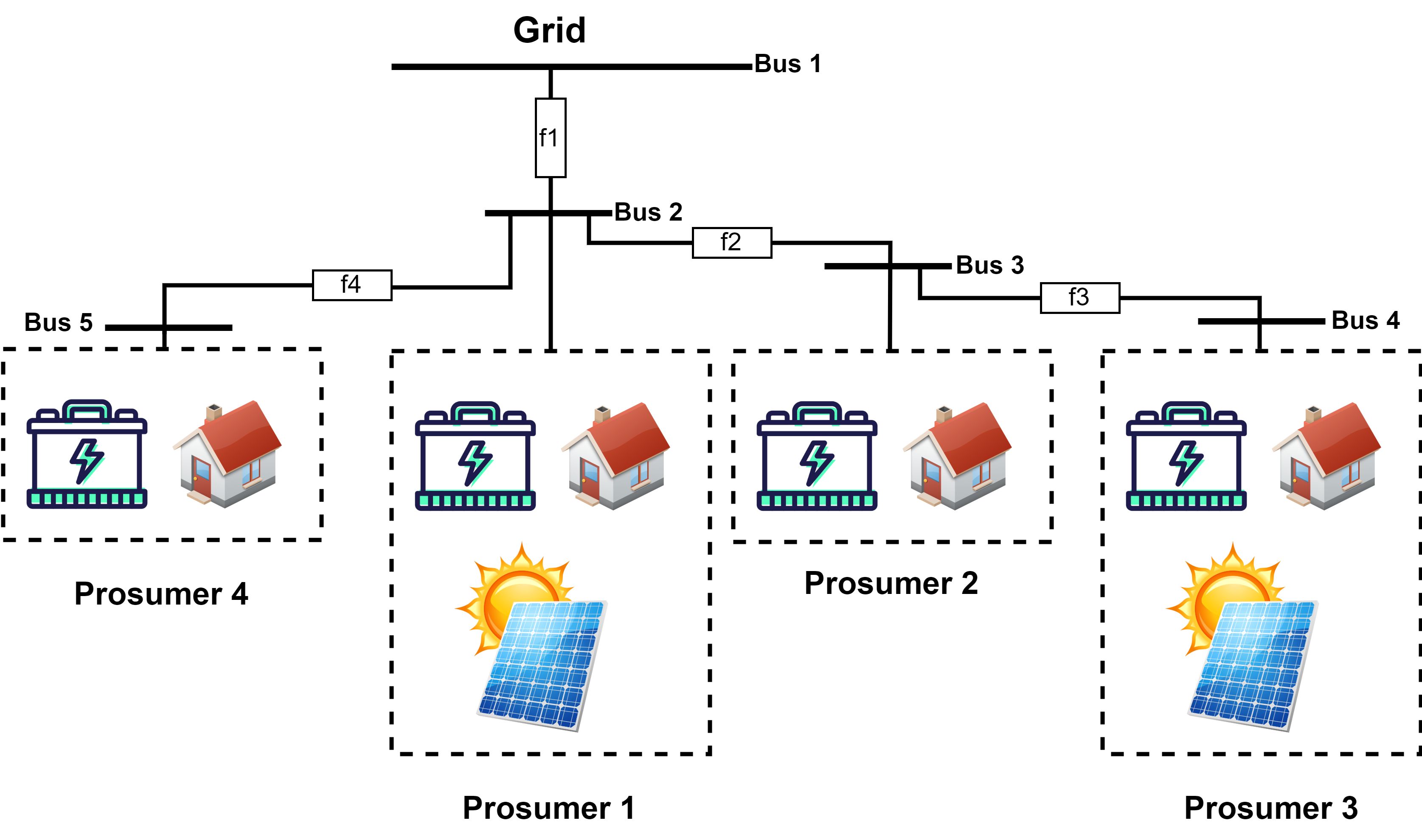}
    \caption{Layout of DC microgrid.}
    \vspace{-3mm}
    \label{fig:dcmicrogrid}
\end{figure}

The DC microgrid under test comprises four PV generation units, four BESS, and four variable DC loads, all interconnected on a common \(400V\) DC bus as shown in Fig. \ref{fig:dcmicrogrid}. Each PV array is interfaced to the DC bus through a DC-DC boost converter that provides bus voltage stabilization and allows power injection. The BESS units employ bidirectional buck-boost converters, enabling both charging (buck mode) and discharging (boost mode) at rates up to 1.5C of their energy capacity. The entire specifications of the DC microgrid are given in Table \ref{tab:specs}. This system is implemented in Simulink and compiled to C code for real‐time execution in OPAL-RT \cite{DCGTG}.

\begin{table}[!t]
  \caption{DC microgrid specifications.}
  \label{tab:specs}
  \centering
  \renewcommand{\arraystretch}{1.2}
  \begin{tabular*}{\columnwidth}{@{\extracolsep{\fill}}cc@{}}
    \toprule
    \textbf{Description} & \textbf{Value and Unit} \\
    \midrule
    \multicolumn{2}{c}{\textbf{PV Ratings}} \\
    \midrule
    PV\,1 at bus\,2   & 1450\,W \\
    PV\,2 at bus\,4   & 450\,W \\
    \midrule
    \multicolumn{2}{c}{\textbf{BESS Ratings}} \\
    \midrule
    BESS at bus\,2    & 1\,kWh \\
    BESS at bus\,3    & 2\,kWh \\
    BESS at bus\,4    & 1\,kWh \\
    BESS at bus\,5    & 2\,kWh \\
    \midrule
    \multicolumn{2}{c}{\textbf{Load Ratings}} \\
    \midrule
    Load at bus\,2    & Max: 160\,W, Min: 130\,W \\
    Load at bus\,3    & Max: 720\,W, Min: 240\,W \\
    Load at bus\,4    & Max: 700\,W, Min: 110\,W \\
    Load at bus\,5    & Max: 1100\,W, Min: 210\,W \\
    \midrule
    \multicolumn{2}{c}{\textbf{Feeder Specifications}} \\
    \midrule
    f1                & $R=1.257\,\Omega$, $L=0.031\,\mathrm H$ \\
    f2                & $R=1.150\,\Omega$, $L=0.030\,\mathrm H$ \\
    f3                & $R=0.868\,\Omega$, $L=0.028\,\mathrm H$ \\
    f4                & $R=0.469\,\Omega$, $L=0.035\,\mathrm H$ \\
    \bottomrule
  \end{tabular*}
\end{table}

The mathematical modeling of the components in the DC microgrid is as follows. The PV converter is modeled as an average model given by (\ref{eq:pv_boost}): 
\begin{align}
L_{p,i}\,\frac{d i_{p,i}}{d t} 
&= v_{pv,i} - d_{p,i}\,v_{\rm dc}, 
\label{eq:pv_boost}\\
0 \;\le\; d_{p,i} &\;\le\; 1,
\end{align}
where \(i_{p,i}\) is the \(ith\) converter inductor current, \(v_{pv,i}\) the panel voltage, \(d_{p,i}\) the duty ratio, and \(L_{p,i}\) the boost converter inductance.

The dynamics of DC‐bus voltage \(v_{\rm dc}\) follows Kirchhoff’s current law and is given by (\ref{eq:dc_bus}):
\begin{equation}
C_{dc}\,\frac{d v_{\rm dc}}{d t}
=\sum_{i=1}^{4}d_{p,i}\,i_{p,i}
 - \sum_{i=1}^{4}i_{b,i}
 - \sum_{i=1}^{4}i_{L,i},
\label{eq:dc_bus}
\end{equation}
where \(C_{dc}\) is the bus capacitance, \(i_{b,i}\) is the BESS converter current, and \(i_{L,i}\) is the load current at bus \(i\).
The BESS is connected to the DC bus through a bidirectional converter which is modeled as in (\ref{eq:bess_conv}):
\begin{align}
L_{b,i}\,\frac{d i_{b,i}}{d t}
&= d_{b,i}\,v_{\rm dc} - v_{b,i}, 
\label{eq:bess_conv}\\
-1 \;\le\; d_{b,i} &\;\le\; 1,
\end{align}
where \(d_{b,i}\) is the converter duty (positive for discharge, negative for charge), \(v_{b,i}\) is the battery terminal voltage, and \(L_{b,i}\) is the converter inductance.  The battery SoC of the BESS at any instant (\(E_i\)) is given by (\ref{eq:soc}):
\begin{equation}
\frac{d E_i}{d t} = -\,\eta_i\,i_{b,i},
\label{eq:soc}
\end{equation}
where \(\eta_i\) is the round‐trip efficiency and \(i_{b,i}\) is the charging/discharging current. The load in the DC microgrid is modeled as a time-varying conductance and is given by (\ref{eq:load}).
\begin{equation}
i_{L,i}(t) = G_{L,i}(t)\,v_{\rm dc}(t).
\label{eq:load}
\end{equation}

\subsection{Real-time CPS Testbed}
The DC microgrid is modeled in Simulink and deployed for real-time execution on the OPAL-RT platform. All components, including PV arrays, BESS units, power converters, and loads, are implemented in a Simulink model that is adapted to run in real-time in OPAL-RT. The DC loads are represented as dynamic resistive networks with programmable power profiles, enabling the simulation of time-varying load conditions. These loads can be reconfigured during runtime through controller-issued commands, allowing the testbed to emulate grid disturbances, load shedding events, and step changes in power demand.

\begin{figure*}[t]
    \centering
    \includegraphics[width=0.95\textwidth]{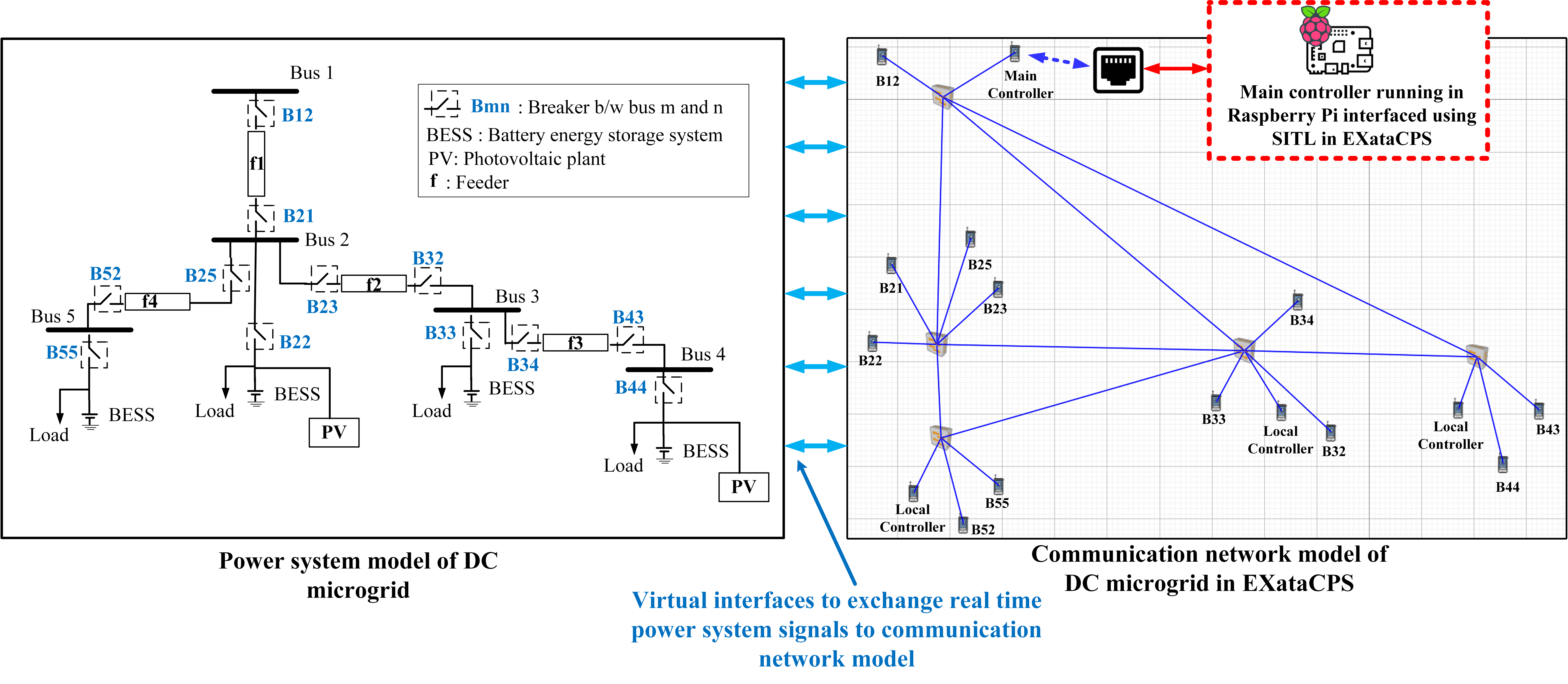} 
    \caption{Cyber-physical setup integrating power system and communication network simulation.}
    \label{fig:cyberphysical}
    \vspace{-3mm}
\end{figure*}

The CHIL setup is implemented using EXataCPS, a real-time network emulation platform designed to model cyber-physical system interactions. EXataCPS is executed on a dedicated core of the OPAL-RT simulator to enable tight synchronization with the power system model. Communication between the cyber and physical domains is established via virtual Ethernet interfaces, allowing real-time, bidirectional exchange of measurements and control signals. Modbus over TCP is selected as the underlying communication protocol.  Circuit breakers on each line, local controllers for individual prosumers, and the main controller are modeled as communication nodes within OPAL-RT and mapped to corresponding nodes in EXataCPS. This configuration redirects traffic from the power system simulation to the emulated communication network. Each node is assigned a unique IP address and communicates using the Modbus TCP protocol. The main controller functions as the Modbus master, while all other nodes act as Modbus slaves. The cyber-physical setup showing integration of power system simulation and communication network simulation is shown in Fig. \ref{fig:cyberphysical}. 

\begin{figure}[t]
  \centering
  \includegraphics[width=0.95\linewidth]{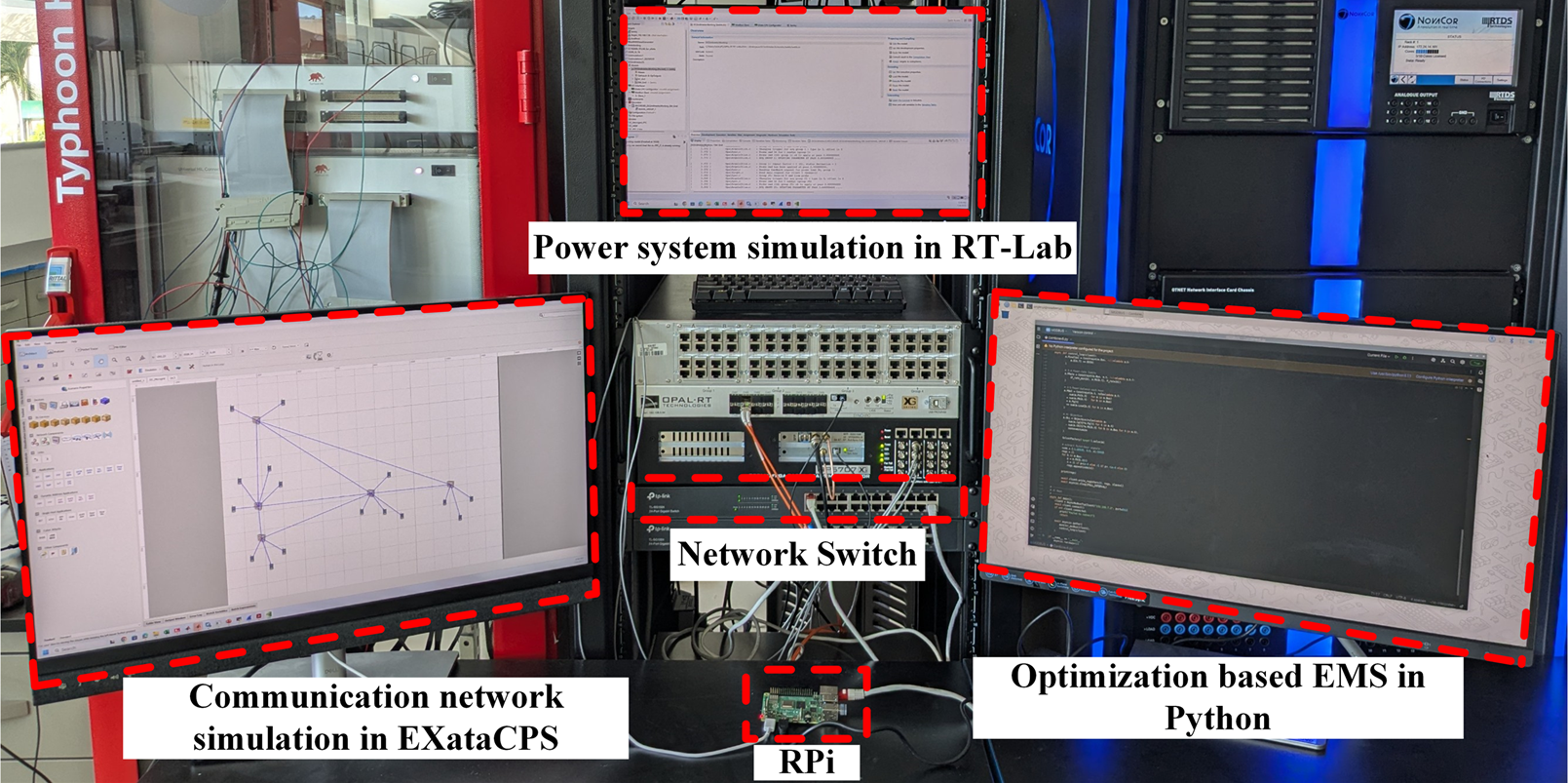}
  \caption{Lab setup of the CPS testbed.}
  \vspace{-3mm}
  \label{fig:labsetup}
\end{figure}

The main controller runs on RPi and is integrated into the setup as an external device through the system-in-the-loop (SITL) in EXataCPS (highlighted in the red dashed box). This is done via adding the MAC address and IP address of the RPi as an Address Resolution Protocol (ARP) entry in the EXataCPS configurator in OPAL-RT. This allows the real Modbus traffic originating on RPi to enter the simulated communication network in EXataCPS. The lab setup of the CPS testbed is shown in Fig. \ref{fig:labsetup}.



The optimization‐based EMS architecture is organized into three functional layers, which are as follows:
\begin{enumerate}
  \item \textbf{Data Acquisition Layer:}  
    Measurement interfaces collect SoC data from each BESS and retrieve PV irradiance and load measurements.
  \item \textbf{Optimization Layer:}  
    Model formulation encapsulates system dynamics, device limits, and economic objectives into a linear program. The solver engine executes the optimization at defined intervals, producing a sequence of control setpoints for grid import/export and BESS charge/discharge.
  \item \textbf{Control \& Communication Layer:}  
    Signal dispatch translates optimized setpoints into discrete control commands (charge, idle, discharge) and writes them to the BESS controllers via Modbus. Supervisory monitoring verifies execution fidelity and triggers re‐optimization upon deviations.
\end{enumerate}

The EMS is deployed on an RPi and communicates with the OPAL-RT simulator via the Modbus/TCP protocol to acquire real-time data, including nodal voltages, currents, SoC values, and load profiles. The EMS adopts a hierarchical two-time-scale control strategy, consisting of a fast monitoring stage and a slow optimization stage.
In the fast monitoring layer (sub-second resolution), the EMS continuously polls SoC and grid measurement registers at millisecond intervals to capture transient phenomena and maintain accurate situational awareness.
In the slow optimization layer (multi-minute resolution), the EMS performs a rolling 24-hour receding-horizon optimization every five minutes. From each optimization run, only the control actions for the immediate next hour are implemented in real-time. The horizon is then advanced to incorporate updated measurements and forecasts for the subsequent optimization cycle.

The EMS is implemented in Python using Pyomo for model declaration and IPOPT for solving. Asynchronous I/O with \texttt{asyncio} library enables millisecond‐scale Modbus polling alongside nonblocking solver execution. Control signals (\(-1,0,+1\)) are encoded into 16‐bit registers to command each BESS in charge, idle, or discharge mode.



\section{Results and Discussion}\label{sec:results}

\begin{figure}[t]
    \centering
    \includegraphics[scale=1]{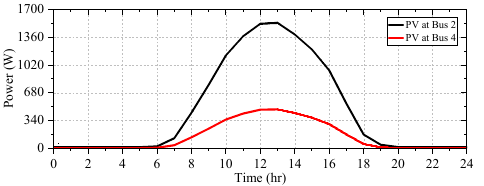}
    \caption{Power output of PV plants in DC microgrid.}
    \label{fig:pv}
\end{figure}

\begin{figure}[t]
    \centering
    \includegraphics[scale=1]{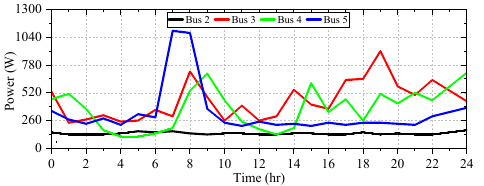}
    \caption{Load profile of individual loads at their respective buses.}
    \label{fig:load}
\end{figure}

\begin{figure}[t]
    \centering
    \includegraphics[scale=1]{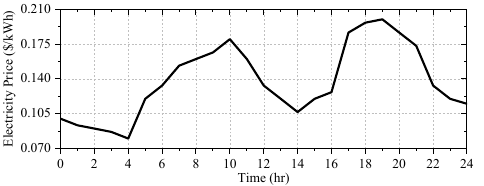}
    \caption{Real-time price profile.}
    \label{fig:price}
\end{figure}

\begin{figure}[t]
    \centering
    \includegraphics[scale=1]{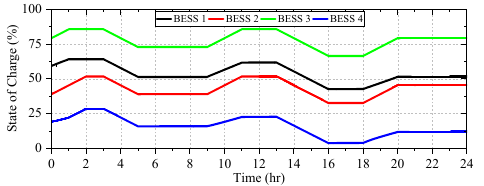}
    \caption{State-of-charge (SoC) of different BESS units with EMS.}
    \label{fig:soc}
\end{figure}

\begin{figure}[t]
    \centering
    \includegraphics[scale=1]{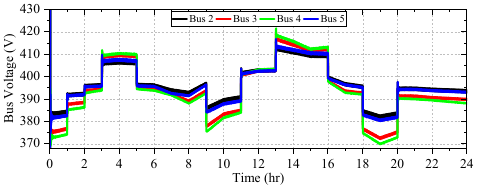}
    \caption{Bus voltages with EMS.}
    \label{fig:vol}
\end{figure}

\begin{figure}[t]
    \centering
    \includegraphics[scale=1]{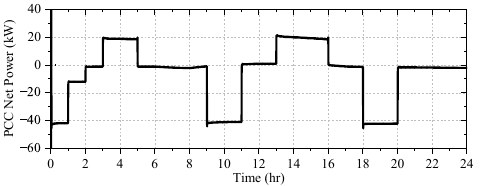}
    \caption{Net power exchange from PCC with EMS.}
    \label{fig:pcc}
\end{figure}

The proposed EMS effectively minimizes operational costs by dynamically optimizing power exchange between the local microgrid, including a BESS, and the main grid in response to real-time electricity price fluctuations. The performance of the EMS is demonstrated through key operational parameters, which are discussed in this section.

Fig. \ref{fig:pv} illustrates the typical daily PV profile, which exhibits a peak during midday hours. However, the intermittency of solar power results in zero generation at night. During these periods of insufficient PV generation, the EMS dispatches power from the BESS or procures it from the main grid to ensure a continuous and uninterrupted power supply to all connected loads, thereby maintaining system reliability while adhering to the cost minimization objective.

The load profiles for all four buses, presented in Fig. \ref{fig:load}, reveal distinct demand patterns characterized by prominent morning and evening peaks. These profiles accurately reflect typical residential and commercial energy consumption behaviors. Fig. \ref{fig:price} depicts the time-varying nature of grid electricity prices \cite{RTP}. Significant fluctuations are observed throughout the day, with noticeably higher rates coinciding with peak demand periods (morning and evening) and lower rates during off-peak hours (late night and early morning). This dynamic pricing structure forms the cornerstone of the EMS's economic optimization strategy. By accurately responding to these price signals, the EMS is able to make informed decisions regarding power procurement and BESS operation.

The efficacy of the EMS's cost-minimization strategy is demonstrated in Fig. \ref{fig:soc}, which tracks the battery's SoC over a 24-hour period. The EMS strategically charges the BESS during periods of low grid electricity prices and subsequently discharges it during high-price periods. By actively managing the SoC, the EMS ensures that the battery operates within a healthy range of 15\% to 95\%, thereby avoiding deep discharging or overcharging, which are detrimental to battery lifespan. This proactive management of SoC underscores a key advantage of the EMS in balancing economic efficiency with battery health preservation.

As shown in Fig. \ref{fig:vol}, the voltage stability at all buses within the microgrid remains well within acceptable operational limits throughout the simulation period. This finding confirms that despite the dynamic power exchanges and optimization strategies employed by the EMS, the power quality within the microgrid is consistently maintained.

Fig. \ref{fig:pcc} illustrates the bidirectional power exchange at the point of common coupling (PCC) between the microgrid and the main grid. Positive values indicate power import, where the microgrid draws energy from the main grid, while negative values represent power export, where excess energy is fed back into the grid. This controlled interaction, managed by the EMS, not only minimizes operational costs but also enhances the stability and reliability of the broader distribution network by mitigating power fluctuations.

The performance evaluation of the proposed EMS under realistic communication delays was carried out in the CPS testbed by analyzing communication delays and jitter across five traffic classes \cite{Aftab2018} (DS0, DS1, DS3, E1, E3) under increasing congestion levels (0\% to 75\%). As shown in Table \ref{tab:avg_delay}, lower-priority traffic (DS0) exhibits significant latency degradation, with average delay rising from 24.25 ms (0\% congestion) to 56.56 ms (75\% congestion). In contrast, higher-priority classes (DS3, E3) maintain stable delays below 2.2 ms even under severe congestion, demonstrating effective quality of service prioritization.

Jitter performance follows a similar trend as seen in Table \ref{tab:avg_jitter}. DS0 experiences a 13 times increase in jitter (135 µs to 1770 µs) at 75\% congestion, while DS3 and E3 remain below 0.35 µs, critical for time-sensitive EMS operations. The divergence in performance between traffic classes underscores the importance of assigning EMS-related data like BESS control signals, bus voltages, currents, and SoC updates to high-priority channels (DS3/E3) to mitigate instability risks.

\begin{table}[!t]
  \caption{Average delay (ms) under different congestion levels.}
  \label{tab:avg_delay}
  \centering
  \renewcommand{\arraystretch}{1.2}
  \begin{tabular*}{\columnwidth}{@{\extracolsep{\fill}}cccccc@{}}
    \toprule
    \textbf{Congestion (\%)} & \textbf{DS0} & \textbf{DS1} & \textbf{DS3} & \textbf{E1} & \textbf{E3} \\
    \midrule
    0   & 24.25 & 2.92   & 2.03   & 2.69 & 2.04   \\
    25  & 27.91 & 3.077  & 2.037  & 2.81 & 2.048  \\
    50  & 35.13 & 3.385  & 2.048  & 3.04 & 2.059  \\
    75  & 56.56 & 4.307  & 2.081  & 3.73 & 2.103  \\
    \bottomrule
  \end{tabular*}
\end{table}

\begin{table}[!t]
  \caption{Average jitter ($\mu$s) under different congestion levels.}
  \label{tab:avg_jitter}
  \centering
  \renewcommand{\arraystretch}{1.2}
  \begin{tabular*}{\columnwidth}{@{\extracolsep{\fill}}cccccc@{}}
    \toprule
    \textbf{Congestion (\%)} & \textbf{DS0} & \textbf{DS1} & \textbf{DS3} & \textbf{E1} & \textbf{E3} \\
    \midrule
    0    & 135   & 1.437   & 0.063   & 1.14  & 0.138 \\
    25   & 152   & 1.905   & 0.0638  & 1.853 & 0.101 \\
    50   & 401   & 3.693   & 0.119   & 2.6   & 0.155 \\
    75   & 1770  & 8.205   & 0.228   & 5.02  & 0.322 \\
    \bottomrule
  \end{tabular*}
\end{table}

\section{Conclusion}\label{sec:conclusion}
This paper presents a comprehensive real-time CPS testbed for evaluating energy management strategies in DC microgrids under realistic communication delays. The testbed integrates a DC microgrid model executed on OPAL-RT, an optimization-based EMS running on an RPi, and a communication network emulated using EXataCPS. By incorporating realistic power system traffic and communication latency, the testbed enables thorough validation of EMS control strategies under practical cyber-physical constraints. This integrated platform facilitates co-validation of control algorithms, communication protocols, and hardware interfaces, providing a robust environment for developing resilient and deployable EMS solutions in DC microgrids.

Results reveal that unprioritized traffic degrades significantly under congestion, underscoring the need for redundancy and strict quality of service enforcement for latency-sensitive EMS functions. Future work will extend this framework to evaluate cybersecurity vulnerabilities and incorporate adaptive control strategies in the presence of network disruptions or malicious events.

\section*{Acknowledgments}
This publication is based upon work supported by King Abdullah University of Science and Technology (KAUST) Center of Excellence for Renewable Energy and Storage Technologies (CREST) under award \#5937.

\bibliographystyle{IEEEtran}
\bibliography{References}

@ARTICLE{RealTimeEMS,
  author={Gao, Fei and Yu, Jiahao and Rogers, Daniel J.},
  journal={IEEE Transactions on Power Electronics}, 
  title={A Discrete-Time Algorithm for Real Time Energy Management in DC Microgrids}, 
  year={2023},
  volume={38},
  number={3},
  pages={2896-2909},
 }

@ARTICLE{CentralizedEMS,
  author={Bhattar, Chandrakant L. and Chaudhari, Madhuri A.},
  journal={IEEE Systems Journal}, 
  title={Centralized Energy Management Scheme for Grid Connected DC Microgrid}, 
  year={2023},
  volume={17},
  number={3},
  pages={3741-3751},
}

@ARTICLE{CMS_DC,
  author={Pannala, Sanjeev and Patari, Niloy and Srivastava, Anurag K. and Padhy, Narayana Prasad},
  journal={IEEE Transactions on Industry Applications}, 
  title={Effective Control and Management Scheme for Isolated and Grid Connected DC Microgrid}, 
  year={2020},
  volume={56},
  number={6},
  pages={6767-6780},
  doi={10.1109/TIA.2020.3015819}}

@ARTICLE{EMS_DC,
  author={Wang, Shuoqi and Du, Mingqiao and Lu, Languang and Xing, Wei and Sun, Kai and Ouyang, Minggao},
  journal={IEEE Journal of Emerging and Selected Topics in Power Electronics}, 
  title={Multilevel Energy Management of a DC Microgrid Based on Virtual-Battery Model Considering Voltage Regulation and Economic Optimization}, 
  year={2021},
  volume={9},
  number={3},
  pages={2881-2895},
  doi={10.1109/JESTPE.2020.2975904}}

@ARTICLE{SoC_DC,
  author={Panda, Mrutunjaya and Chankaya, Mukul and Mohanty, Satyajit and Sandeep, S. D.},
  journal={IEEE Access}, 
  title={State-of-Charge-Based Energy Management Strategy for Hybrid Energy Storage System in DC Microgrid}, 
  year={2025},
  volume={13},
  number={},
  pages={77353-77364},
  doi={10.1109/ACCESS.2025.3564786},}

@ARTICLE{OPF_undercommdelay,
  author={Hu, Jian and Ma, Hao},
  journal={Journal of Modern Power Systems and Clean Energy}, 
  title={Distributed Real-time Optimal Power Flow Strategy for DC Microgrid Under Stochastic Communication Networks}, 
  year={2023},
  volume={11},
  number={5},
  pages={1585-1595},
  doi={10.35833/MPCE.2021.000730},}

@article{DER_Review,
title = {Distributed Energy Resources: A Systematic Literature Review},
journal = {Energy Reports},
volume = {13},
pages = {1980-1999},
year = {2025},
issn = {2352-4847},
doi = {https://doi.org/10.1016/j.egyr.2025.01.026},

author = {Midrar Adham and Sean Keene and Robert B. Bass},
}

@ARTICLE{DCuG_Review,
  author={Dragičević, Tomislav and Lu, Xiaonan and Vasquez, Juan C. and Guerrero, Josep M.},
  journal={IEEE Transactions on Power Electronics}, 
  title={DC Microgrids—Part II: A Review of Power Architectures, Applications, and Standardization Issues}, 
  year={2016},
  volume={31},
  number={5},
  pages={3528-3549},
  doi={10.1109/TPEL.2015.2464277},}

@article{EMSinMG,
title = {An Energy Management System for Multi-Microgrid system considering uncertainties using multi-objective multi-verse optimization},
journal = {Energy Reports},
volume = {13},
pages = {286-302},
year = {2025},
issn = {2352-4847},
doi = {https://doi.org/10.1016/j.egyr.2024.12.001},

author = {Dessalegn Bitew Aeggegn and George Nyauma Nyakoe and Cyrus Wekesa},
}

@ARTICLE{EMS_advantageforRES,
  author={Thirugnanam, Kannan and El Moursi, Mohamed Shawky and Khadkikar, Vinod and Zeineldin, Hatem H. and Hosani, Mohamed Al},
  journal={IEEE Transactions on Smart Grid}, 
  title={Energy Management Strategy of a Reconfigurable Grid-Tied Hybrid AC/DC Microgrid for Commercial Building Applications}, 
  year={2022},
  volume={13},
  number={3},
  pages={1720-1738},
  doi={10.1109/TSG.2022.3141459},}

@article{EMS_advanatgeforRES1,
title = {Economic optimal load management control of microgrid system using energy storage system},
journal = {Journal of Energy Storage},
volume = {46},
pages = {103843},
year = {2022},
issn = {2352-152X},
doi = {https://doi.org/10.1016/j.est.2021.103843},

author = {N.T. Mbungu and T. Madiba and R.C. Bansal and M. Bettayeb and R.M. Naidoo and M.W. Siti and T. Adefarati},
}

@INPROCEEDINGS{DCGTG,
  author={S. Gokul Krishnan and Konstantinou, Charalambos},
  booktitle={2024 12th Workshop on Modeling and Simulation of Cyber-Physical Energy Systems (MSCPES)}, 
  title={Design and Evaluation of a DC Microgrid Testbed for DER Integration and Power Management}, 
  year={2024},
  volume={},
  number={},
  pages={1-6},
  doi={10.1109/MSCPES62135.2024.10542752}}

@article{Aftab2018,
  title = {Performance evaluation of IEC 61850 GOOSE‐based inter‐substation communication for accelerated distance protection scheme},
  volume = {12},
  ISSN = {1751-8695},
  number = {18},
  journal = {IET Generation,  Transmission and; Distribution},
  publisher = {Institution of Engineering and Technology (IET)},
  author = {Aftab,  Mohd Asim and Roostaee,  Saeed and Suhail Hussain,  S.M. and Ali,  Ikbal and Thomas,  Mini S. and Mehfuz,  Shabana},
  year = {2018},
  pages = {4089–4098}
}

@inproceedings{mohammed2024decentralized,
  title={Decentralized Bus Voltage Restoration for DC Microgrids},
  author={Mohammed, Nabil and Ahmed, Shehab and Konstantinou, Charalambos},
  booktitle={2024 9th IEEE Workshop on the Electronic Grid (eGRID)},
  pages={1--6},
  year={2024},
  organization={IEEE}
}

@article{RTP,
title = {Demand flexibility in hydrogen production by incorporating electrical and physical parameters},
journal = {Electric Power Systems Research},
volume = {239},
pages = {111213},
year = {2025},
issn = {0378-7796},

author = {Mohd Asim Aftab and Vipin Chandra Pandey and S. Gokul Krishnan and Faraz Mir and Gerrit Rolofs and Emeka Chukwureh and Shehab Ahmed and Charalambos Konstantinou},

}

\end{document}